\documentclass[acus]{JAC2003}
\usepackage{textcomp,gensymb}
\usepackage{units}
\usepackage{graphicx}
\usepackage{fancyheadings}
\usepackage[amsmath]{benc}
\begin{document}

\title{PHOTONIC CRYSTAL LASER ACCELERATOR STRUCTURES}
\author{B. Cowan\thanks{benc@slac.stanford.edu}, M. Javanmard,
  R. Siemann, SLAC, Stanford, CA 94309, USA}

\maketitle

\thispagestyle{fancy}
\setlength{\headrulewidth}{0pt}
\rhead{SLAC--PUB--9859 \\ ARDB--333 \\ May 2003}

\begin{abstract}
Photonic crystals have great potential for use as laser-driven
accelerator structures.  A photonic crystal is a dielectric structure
arranged in a periodic geometry.  Like a crystalline solid with its
electronic band structure, the modes of a photonic crystal lie in a
set of allowed photonic bands.  Similarly, it is possible for a photonic
crystal to exhibit one or more photonic band gaps, with frequencies in
the gap unable to propagate in the crystal.  Thus photonic crystals
can confine an optical mode in an all-dielectric structure,
eliminating the need for metals and their characteristic losses at
optical frequencies.

We discuss several geometries of photonic crystal accelerator
structures.  Photonic crystal fibers (PCFs) are optical fibers which
can confine a speed-of-light optical mode in vacuum.  Planar
structures, both two- and three-dimensional, can also confine such a
mode, and have the additional advantage that they can be manufactured
using common microfabrication techniques such as those used for
integrated circuits.  This allows for a variety of possible materials,
so that dielectrics with desirable optical and radiation-hardness
properties can be chosen.  We discuss examples of simulated photonic crystal
structures to demonstrate the scaling laws and trade-offs involved,
and touch on potential fabrication processes.

\begin{center}
\emph{Submitted to Particale Accelerator Conference (PAC 2003),
May 12--16, 2003, Portland, Oregon (IEEE)}
\end{center}
\end{abstract}

\section{INTRODUCTION}
The extraordinary electric fields available from laser systems make
laser-driven charged particle acceleration an exciting possibility.
Practical vacuum laser acceleration requires a guided-mode structure
capable of confining a speed-of-light (SOL) mode and composed entirely
of dielectric materials, and photonic crystals provide a means to
achieve this capability.  A photonic crystal is a structure with
permittivity periodic in one or more of its dimensions.  As described
in \cite{Molding}, optical modes in a photonic crystal form bands,
just as electronic states do in a crystalline solid.  Similarly, a
photonic crystal can also exhibit one or more photonic band gaps
(PBG's), with frequencies in the gap unable to propagate in the
crystal.  Confined modes can be obtained by introducing a defect into
a Photonic Crystal lattice.  Since frequencies in the bandgap are
forbidden to propagate in the crystal, they are confined to the
defect.  A linear defect thus functions as a waveguide.

A significant benefit of photonic crystal accelerators is that only
frequencies within a bandgap are confined.  In general, higher order
modes, which can be excited by the electron beam, escape through the
lattice.  This benefit has motivated work on matallic PBG structures
at RF frequencies \cite{MetalPBG}.  In addition, an accelerating mode
has been found in a PBG fiber structure \cite{Eddie}.  After
discussing 2D planar structures we consider the fiber geometry in more
generality.

\section{2D PLANAR PHOTONIC CRYSTAL ACCELERATOR STRUCTURES}
\label{sec:2D}

\subsection{Structure Geometry}

The geometries we consider in this section are two-dimensional: we
take them to be infinite in the vertical ($y$) direction, while the
electron beam and the accelerating optical field copropagate in the
$z$-direction, transverse to the direction of symmetry.  While such
structures are not immediately suitable for charged particle
acceleration, 2D structures can be analyzed with much less CPU time
than can 3D structures, thereby allowing rapid exploration of multiple
sets of geometric parameters.  The computational technique is
discussed further below.

Our underlying photonic crystal lattice is a triangular array of
vacuum holes in a silicon substrate.  Assuming an operating wavelength
of \unit[1.5]{\micro m} in the telecom band, silicon has a normalized
permittivity of $\epsilon_r = \epsilon/\epsilon_0 = 12.1$
\cite{ref:HOCSilicon}.  Such a lattice exhibits a wide TE bandgap, as
desired since the accelerating field component is transverse to the
direction of symmetry.  For lattice constant $a$ the nearest-neighbor
center-to-center hole spacing, we choose the hole radius $r = 0.427a$
to maximize the relative width of the bandgap.

\begin{figure}
\begin{center}
\includegraphics{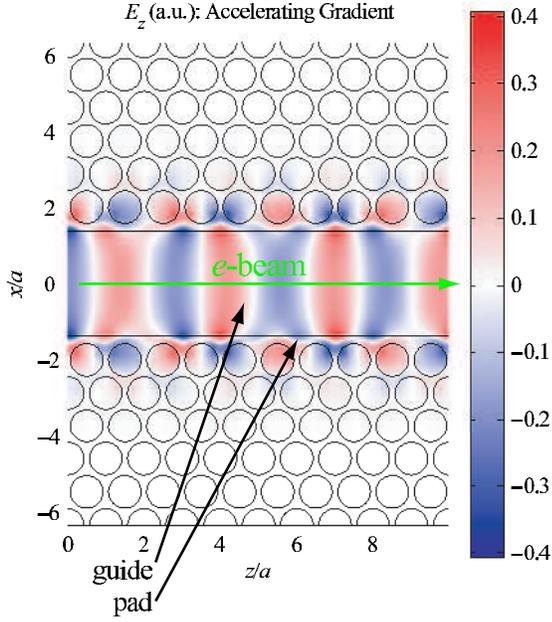}
\caption{An accelerator structure geometry with a waveguide mode.
The shading indicates the electric field component in the direction of
\emph{e}-beam propagation.  Here the guide width is $w = 3.0a$, the
pad width is $\delta = 0.25a$, and the wavelength is $\lambda = 2.78a$.}
\label{fig:mode}
\end{center}
\end{figure}

The accelerator structure consists of a vacuum guide in this lattice,
as shown in Fig.~\ref{fig:mode}.
The guide width $w$ is defined such that the distance between the
centers of the holes adjacent to the waveguide are $w+a$.  Also,
dielectric material can be added to the sides of the guide,
and we let $\delta$ denote the total width of the dielectric ``pad''
added to both sides of the guide.  Fig.~\ref{fig:mode} also shows an
accelerating mode of this geometry, i.e. $E_z$ is nonzero on axis and
$\omega = ck_z$.  In fact, for a general selection of $w$ and
$\delta$, there will be a $k_z$ for which this waveguide mode is
synchronous.  This is because the dispersion properties of this PBG
waveguide are similar to a metallic guide in that $\omega/k_z > c$
throughout most of the bandgap, but at the upper edge of the gap the
dispersion curve reduces in slope and meets the SOL line.  The padding
can be added in order to bring the SOL frequency into the center of
the gap where the dispersion curve is more linear.

\subsection{Accelerating Mode Parameters}

Several parameters characterize the performance of an accelerating
mode.  The relationship between the input laser power and the
accelerating gradient is described by the characteristic impedance
\cite{optacc:Scaling}.  Since our 2D structures only confine modes in
one transverse dimension, we normalize the impedance to that of a
structure one wavelength high, so
$Z_c = E\tsub{acc}^2\lambda/P_h$,
where $E\tsub{acc}$ is the accelerating gradient and $P_h$ is the
laser power per unit height.  We find an empirical power-law scaling
of the impedance, with $Z_c\propto (w/\lambda)^{-3.55}$.

Next, there is the \emph{damage factor} $f_D =
E\tsub{acc}/\abs{\vect{E}}\tsup{mat}\tsub{max}$, where
$\abs{\vect{E}}\tsup{mat}\tsub{max}$ is the maximum electric field
magnitude anywhere in the dielectric material.  Since laser power is
ultimately limited by the damage threshold of the material, the damage
factor is an important measure of the maximum possible accelerating
gradient a structure can sustain.

\begin{figure}
\begin{center}
\resizebox{\linewidth}{!}{\includegraphics{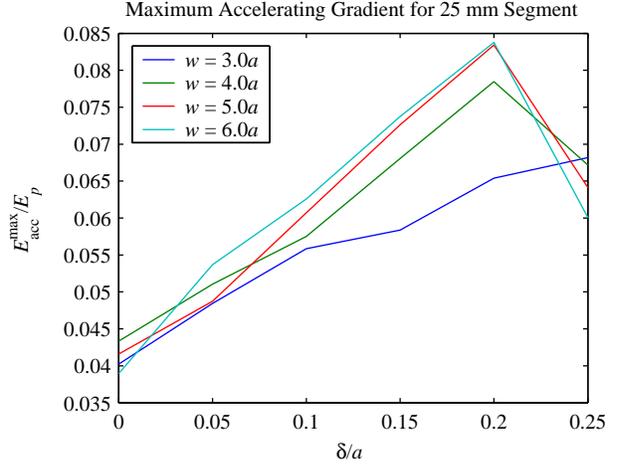}}
\caption{The maximum accelerating gradient sustainable by each
structure geometry, normalized to the material damage threshold $E_p$
for \unit[1]{ps} pulses.}
\label{fig:maxgradient}
\end{center}
\end{figure}

The damage threshold exhibits a dependence on laser
pulse width which becomes favorable at very short pulse widths, as
examined in \cite{Damage} and paramaterized in \cite{optacc:Scaling}.
Thus these accelerator structures are transmission-mode, and a high
group velocity $v_g$ is desired so that short pulses may be used.  The
qualitative behavior of these parameters presents a trade-off.  As the
guide is widened, the damage factor decreases.  On the other hand, the
group velocity increases, allowing shorter laser pulses to be used,
for which the material damage threshold is at a higher field.  To find
the optimum parameters we plot the maximum possible accelerating
gradient taking both effects into account in Fig.~\ref{fig:maxgradient}.

\section{PHOTONIC CRYSTAL FIBER STRUCTURES}

\begin{figure*}
\begin{center}
\includegraphics*[1.228in,4.6in][7.228in,6.4in]{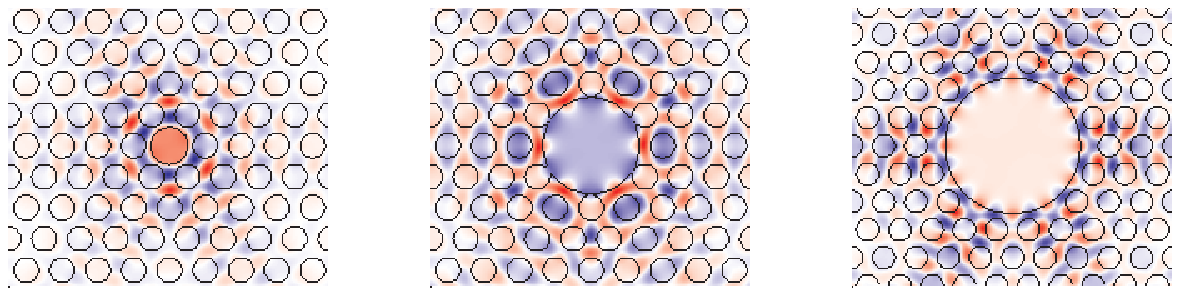}
\vskip -1.8in
\includegraphics*[1.228in,4.6in][7.228in,6.4in]{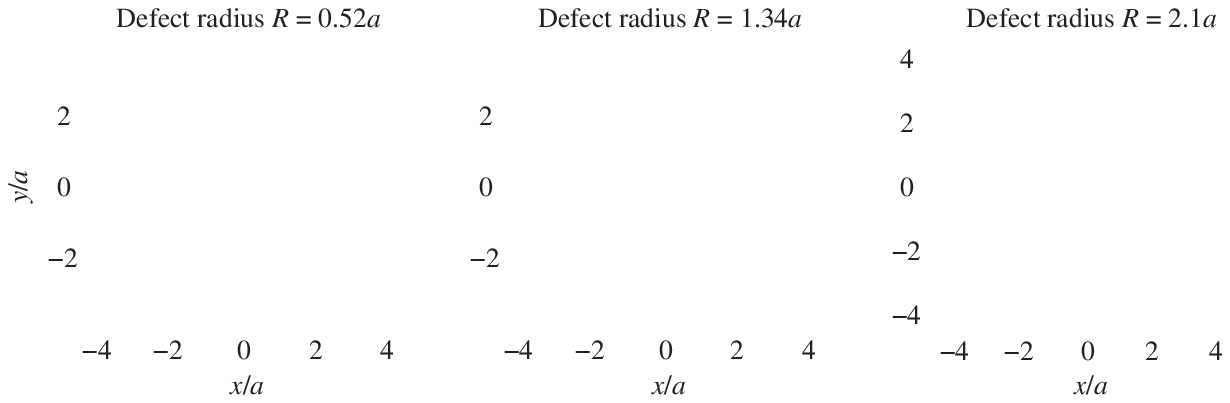}
\caption{Accelerating modes of several photonic crystal fiber
configurations.  The shading indicates the axial electric field, which
is also the direction of propagation of the mode.}
\label{fig:fibermode}
\end{center}
\end{figure*}

The geometry of this structure is again a triangular array of vacuum
holes, this time in silica ($\epsilon_r = 2.13$).  Here the structure
is considered to be a fiber drawn infinitely in the beam propagation
direction, with the electrons and laser pulse copropagating along the
fiber.  For these structures $r = 0.35a$, and the defect consists of a
larger central hole.  Modes were found for three different mode radii,
and are shown in
Fig.~\ref{fig:fibermode}.
The frequencies of the three modes are given by $\omega a/c$ of 8.20,
8.12, and 8.20 and group velocities $0.60c$, $0.654c$, and $0.59c$
(left to right).

The structure was simulated using periodic boundary conditions, and
the fields in the lattice region are due to crosstalk between
neighboring defects.  By increasing the size of the supercell, this
crosstalk can be minimized, however the computational time
significantly increases.  Also, the 6-fold azimuthal symmetry of the
structure implies that SOL modes in vacuum contain only $m=6n$
azimuthal modes for $n$ an integer, reducing the emittance blowup from
higher-order modes.  Finally, we find that the characteristic
impedance decreases with guide radius, as is the case with metallic
waveguide structures.

\section{COMPUTATION TECHNIQUES}
\label{sec:computation}

We use the MIT Photonic Bands (MPB) package, a public-domain code
using an iterative eigensolver technique \cite{MPB}.  For a given
geometry and Bloch wavevector, MPB computes frequencies and field
configurations of supported modes.  MPB can compute group velocities
of modes as well by applying the Feynman-Hellmann theorem
\cite{Sakurai,Perturbation}.

Using the frequencies and group velocities, we can inductively
converge on the longitudinal wavenumber for which a speed-of-light
mode exists.  Having found a mode for a particular wavenumber, we can
use its frequency and group velocity to obtain a linear approximation
for its dispersion curve near that wavenumber.  The intersection of
that approximation with the SOL line gives the wavenumber for the next
computation, which yields a mode whose phase velocity is closer to
$c$.  Since the iterative eigensolver for each step can be seeded with
the field configuration from the result of the previous step,
successive steps are quite fast, and convergence to an SOL mode is a
computationally light task once the initial mode has been found.

\section{FABRICATION POSSIBILITIES}

The 2D structures discussed above are amenable to
photolithography, with 50 : 1 apect ratios available from current
reactive ion etching equipment.  Some investigation into fabrication
of these structures has taken place in the past \cite{Holes}.
Fabrication of 3D photonic crystals with omnidirectional bandgaps,
such as the ``woodpile'' structure \cite{Woodpile}, is an active area
of research.  A number of techniques are being developed, including
multilayer lithography, wafer fusion, stacking by micromanipulation,
self-assembly, and others \cite{Clever}.
PCF manufacturing is a large and growing area of research in industry,
since photonic crystals allow for tailoring optical properties to
specific applications, from nonlinearity for wavelength conversion in
telecommunications to large mode area for materials processing
\cite{Holey}.

\section{CONCLUSION}

Photonic crystals have great promise as potential laser accelerator
structures.  Not only do they support accelerating modes, but
such modes exist for a wide range of geometric parameters.  While the
basic accelerator parameters have been examined, much remains to be
done to understand the properties of these structures.  Wakefield
computations as well as coupling structure design have yet to be
explored.  In addition, there are many photonic crystal lattices for
which accelerating modes have not been computed, including 3D
geometries.  However, manufacturing technology, numerical
simulation capability, and theoretical understanding continue to
progress at an extraordinary rate, driven by industry forces.  We
therefore expect a bright future for photonic crystals as an
accelerator technology.



\begin{thebibliography}{99} 

\bibitem{Molding}
J. D. Joannopoulos, R. D. Meade, and J. N. Winn, \emph{Photonic
Crystals: Molding the Flow of Light} (Princeton University Press,
Princeton, NJ, 1995).
\bibitem{MetalPBG}
M. A. Shapiro et al., Phys. Rev. ST Accel. Beams \textbf{4}, 042001
(2001).
\bibitem{Eddie}
X. E. Lin, Phys. Rev. ST Accel. Beams \textbf{4}, 051301 (2001).
\bibitem{ref:HOCSilicon}
D. F. Edwards, in \emph{Handbook of Optical Constants}, edited by
E.~D.~Palik (Academic Press, 1985), vol.~1, p.~547.
\bibitem{optacc:Scaling}
L. Sch\"{a}chter, R. L. Byer, and R. H. Siemann, in \emph{Advanced
Accelerator Concepts: Tenth Workshop, Mandalay Beach, CA, 2002},
edited by C.~E.~Clayton and P.~Muggli, U.S. Department of Energy
(American Institute of Physics, Melville, NY, 2002), no.~647 in AIP
Conference Proceedings, pp.~310--323.
\bibitem{Damage}
B. C. Stuart et al., Phys. Rev. Lett. \textbf{74},
2248 (1995).
\bibitem{MPB}
S. G. Johnson and J. D. Joannopoulos, Optics Express \textbf{8}, 173 (2001).
\bibitem{Sakurai}
See for instance J. J. Sakurai, \emph{Modern Quantum Mechanics},
Rev. ed. (Addison-Wesley, Reading, MA, 1995)
\bibitem{Perturbation}
S. G. Johnson et al., Phys. Rev. E \textbf{65}, 066611 (2002).
\bibitem{Holes}
Wendt et al., J. Vac. Sci. Technol. B \textbf{11}, 2637 (1993).
\bibitem{Woodpile}
S. Y. Lin et al., Nature \textbf{394}, 251 (1998).
\bibitem{Clever}
S. G. Johnson, \emph{Fabrication of Three-Dimensional Crystals: Those
Clever Experimentalists}, from lecture series \emph{Photonic Crystals:
Periodic Surprises in Electromagnetism},
\texttt{http://ab-initio.mit.edu/photons/tutorial/}
\bibitem{Holey}
Ren\'{e} Engel Kristiansen, \emph{Guiding Light with Holey Fibers}, OE
Magazine June 2002, p. 25.

\end{thebibliography}
\end{document}